\documentclass[12pt,color]{article}
\usepackage{epsf}
\usepackage{color}
\usepackage{a4}
\usepackage{amsfonts}		% open math symbols
\usepackage{cite}		% collapse citations
\usepackage{color}
\usepackage{multirow}
\usepackage{amsmath}
\usepackage{amssymb}

\def\hybrid{\topmargin 0pt
        \oddsidemargin 0pt
        \headheight 0pt \headsep 0pt
        \textwidth 6.25in       % A4 paper
        \textheight 9.5in       % A4 paper
        \marginparwidth .875in
        \parskip 5pt plus 1pt   \jot = 1.5ex}

\catcode`\@=11
\def\marginnote#1{}

\newcount\hour
\newcount\minute
\newtoks\amorpm
\hour=\time\divide\hour by60
\minute=\time{\multiply\hour by60 \global\advance\minute by-\hour}
\edef\standardtime{{\ifnum\hour<12 \global\amorpm={am}%
        \else\global\amorpm={pm}\advance\hour by-12 \fi
        \ifnum\hour=0 \hour=12 \fi
        \number\hour:\ifnum\minute<10 0\fi\number\minute\the\amorpm}}
\edef\militarytime{\number\hour:\ifnum\minute<10 0\fi\number\minute}

\def\draftlabel#1{{\@bsphack\if@filesw {\let\thepage\relax
   \xdef\@gtempa{\write\@auxout{\string
      \newlabel{#1}{{\@currentlabel}{\thepage}}}}}\@gtempa
   \if@nobreak \ifvmode\nobreak\fi\fi\fi\@esphack}
        \gdef\@eqnlabel{#1}}
\def\@eqnlabel{}
\def\@vacuum{}
\def\draftmarginnote#1{\marginpar{\raggedright\scriptsize\tt#1}}

\def\draft{\oddsidemargin -.5truein
        \def\@oddfoot{\sl preliminary draft \hfil
        \rm\thepage\hfil\sl\today\quad\militarytime}
        \let\@evenfoot\@oddfoot \overfullrule 3pt
        \let\label=\draftlabel
        \let\marginnote=\draftmarginnote
   \def\@eqnnum{(\theequation)\rlap{\kern\marginparsep\tt\@eqnlabel}%
\global\let\@eqnlabel\@vacuum}  }

\def\numberbysection{\@addtoreset{equation}{section}
        \def\theequation{\thesection.\arabic{equation}}}

\def\titlepage{\@restonecolfalse\if@twocolumn\@restonecoltrue\onecolumn
     \else \newpage \fi \thispagestyle{empty}\c@page\z@
        \def\thefootnote{\fnsymbol{footnote}}
	\setcounter{page}{0} }
\def\endtitlepage{\if@restonecol\twocolumn \else  \fi
        \def\thefootnote{\arabic{footnote}}
        \setcounter{footnote}{0}}  %\c@footnote\z@ }
\definecolor{c1}{rgb}{1, 0, 0}
\definecolor{c2}{rgb}{0, 1, 0}
\definecolor{c3}{rgb}{0, 0, 1}
\definecolor{c4}{rgb}{1, 0, 1}
\definecolor{c5}{rgb}{0, 1, 1}

\catcode`@=12
\relax

\def\beq{\begin{equation}}
\def\eeq{\end{equation}}
\def\bea{\begin{eqnarray}}
\def\eea{\end{eqnarray}}
\relax
\numberbysection
\hybrid
\begin{document}
%\begin{titlepage}
\begin{center}
{\large\bf
Persistence in the two dimensional ferromagnetic Ising model
}\\[.3in] 

{\bf T.~Blanchard, L. F. Cugliandolo and M. Picco}\\

%\textcolor{red}{
%\textit{Sorbonne Universit\'es, UPMC Univ Paris 06, UMR 7589, LPTHE, F-75005,
%Paris, France}\\
%\textit{CNRS, UMR 7589, LPTHE, F-75005, Paris, France}
%}
%{
\it Sorbonne Universit\'es, Universit\'e Pierre et Marie Curie - Paris 06, CNRS UMR 7589, 
Laboratoire de Physique Th\'eorique et Hautes Energies, \\
75252 Paris Cedex 05, France
%}
         \\
%         {\it  Bo\^{\i}te 126, Tour 13-14, 5 \`eme \'etage, \\
%               4 place Jussieu,
%               F-75252 Paris CEDEX 05, France \\ }
{\it     e-mail: {\tt blanchard,leticia,picco@lpthe.jussieu.fr} }\\
\end{center}
%\vskip .04in
\centerline{(Dated: \today)}
\vskip .2in
\centerline{\bf ABSTRACT}
\begin{quotation}
We present very accurate numerical estimates of the time and size dependence of the zero-temperature local persistence in the $2d$ ferromagnetic Ising model. 
We show that the effective exponent 
decays algebraically to an asymptotic value $\theta$ that depends upon 
the initial condition. More precisely, 
we find that $\theta$ takes one universal value $0.199(2)$ for initial conditions with short-range spatial correlations as  in a paramagnetic state, 
and the value $0.033(1)$ for initial conditions with the long-range spatial correlations of the  critical Ising state. We checked universality by working 
with a square and a triangular lattice, and by imposing free and periodic boundary conditions. We found that the effective exponent suffers from 
stronger finite size effects in the former case.
\vskip 0.5cm 
\noindent
%\pacs
%PACS numbers: 05.70.Jk, 64.60.Ak, 64.60.Fr
% PACS 1999 classification: http://www.aip.org/pacs/pacs99/pacscheme.html

\end{quotation}
%\end{titlepage}
\section{Introduction}
\label{sec:introduction}

The {\it persistence} is a quantitative measure of the memory of a reference state, typically chosen to be the 
initial condition after an instantaneous quench, that a stochastic process
keeps  along its evolution. It is a very general notion related to first passage times. Research on this 
topic has been intense in the last two decades and it has been thoroughly summarised in a recent  review article~\cite{BMS}. 

For an unbiased  random walker on a line that starts from a positive position at the initial  time $t=0$,  the persistence is the probability that its 
position remains positive up to time $t$~\cite{math}.  For spin systems the zero-temperature local persistence simply
equals the fraction of spins that have never flipped at time $t$ since a zero-temperature quench performed
at time $t=0$. Equivalently, it is  the probability that a single spin has never flipped between the initial time $t=0$ and  
time $t$ under these conditions. Many other physical and mathematical problems where the persistence ideas can be explored are described in~\cite{BMS}. It is a particularly interesting quantity 
as it  depends on the whole history of  the system.

In many relevant extended systems the persistence probability, or persistence in short, decreases algebraically at long times
\begin{equation}
P(t, L\to\infty)\sim t^{-\theta}
\end{equation}
(in the infinite size limit) with $\theta$ a non-trivial  dynamical exponent. 

Derrida, Bray and Godr\`eche defined and computed  $P(t,L)$ numerically in the zero-temperature Glauber Ising chain 
evolved from a random initial condition~\cite{DBG0}. They later 
calculated $P(t,L\to\infty)$ in the  one-dimensional time-dependent Ginzburg-Landau equation with an initial condition with 
a finite density of domain walls~\cite{DBG}. 
More generally,  the persistence was evaluated in the  $d$-dimensional $Q$-state Potts model's Glauber evolution
after a zero-temperature quench from infinite temperature~\cite{DBG0,Stauffer}. The exponent $\theta$ turns out to be  
$Q$ and $d$ dependent. The
exact value of $\theta$ was obtained explicitly in $d=1$ for all values of $Q$ by mapping the domain walls to particles
which diffuse and annihilate after collision~\cite{derrida_exact_1995,derrida_exact_1996}. No such exact results exist in
higher dimensions and one has to rely on numerical estimates of $\theta$. In~\cite{DBG0}, an estimate $\theta = 0.22 (3) $ was obtained for the  
Ising model in dimension two and this result 
was confirmed in~\cite{Stauffer} where the dimensions one to five were considered. 
 These first numerical estimates were followed by the
analytic value $\theta=0.19$ obtained by Majumdar and Sire~\cite{MS,sire_analytical_2000} 
with a perturbation scheme around the 
Gaussian and Markovian stochastic process that mimics curvature driven domain growth~\cite{Ohta}.
In~\cite{yurke_experimental_1997} Yurke \textit{et al.} measured  $\theta=0.19(3)$ in a $2d$
liquid crystal sample in the same universality class as the $2d$IM with non-conserved order parameter dynamics. Other
numerical estimates are $\theta = 0.21$~\cite{manoj_persistence_2000}, $\theta = 0.209(4)$~\cite{jain_zero-temperature_1999}
and $\theta = 0.22$~\cite{Drouffe-Godreche}. These
various estimates are roughly compatible with each other. Still, as we will explain in the main text, they are not fully satisfactory and there is room 
for improvement in the determination of $\theta$.

The focus of our paper is the precise numerical evaluation of the $\theta$ exponent in the $2d$ Ising model with zero-temperature dynamics 
that do not conserve the local magnetization.
As detailed in the previous paragraph, the persistence in this problem 
has been mostly studied by using infinite temperature, \textit{i.e.} uncorrelated, initial
conditions. We will first present a more accurate numerical estimate of $\theta$ for this kind of initial states, and we will later analyze the
persistence for critical Ising, that is to say long-range correlated, initial conditions. We will study the model on square and triangular lattices with 
free and periodic boundary conditions to check for universality. We will pay special attention to finite time and finite size effects.

The paper is organised as follows. In Sec.~\ref{sec:model}
we define the model and numerical method. We also discuss the system sizes and time-scales to be used numerically. 
Section~\ref{sec:persistence} is devoted to the presentation of our results. In Sec.~\ref{sec:outlook} we discuss some 
lines for future analytic and numeric research on persistence in spin models.
 
 \section{Model and simulations}
\label{sec:model}

We consider the ferromagnetic finite dimensional Ising model 
\begin{equation}
H=-J \sum_{\langle ij\rangle} S_i S_j,
\label{H}
\end{equation}
where the  spin variables $S_i=\pm 1$ sit on each site of a two dimensional lattice, 
the sum over $\langle ij \rangle$ is restricted to the nearest neighbours on the lattice,
and $J$ is a positive parameter that we  fix to take the value $J=1$. 
With this choice, the model undergoes a second order phase transition at 
$\beta^{sq} = 1/(k_BT) = {1\over 2} \log{(1+\sqrt{2})}$ for the square
lattice and $\beta^{tr} = 1/(k_BT) = \log{(1+\sqrt{3})}$ for the triangular lattice. 
We will consider both types of lattices with $N=L\times L$ spins and either  
free boundary conditions (FBC) or  periodic boundary conditions (PBC). 
The choice of boundary conditions can have an influence on the final state after a quench
to zero-temperature\cite{SKR} --\cite{BCCP}.  More precisely, the dynamics at low temperature are dominated by 
the coarsening of domains with a linear length scale  that  grows in time as $t^{1/z}$ with the dynamic exponent $z=2$ for non-conserved
order parameter dynamics~\cite{Bray,Malte}, the kind of evolution we use here. Thus, there is a characteristic 
equilibration time, $t_{eq} \simeq L^2$, in this problem. After this time, most of the finite domains have disappeared 
and the configuration is either completely magnetised, such that all the spins take the same value, or in a striped state 
with interfaces crossing the lattice. Depending on the geometry of the  lattice and the choice of boundary conditions, 
these striped states can be stable or not. In the latter case, there is some additional evolution taking place in  a longer time scale. 
In particular, for PBC on the square lattice, there exist diagonal stripe states with a characteristic time 
$t^d_{eq} \simeq L^{3.5}$~\cite{SKR2}, while these do not exist for FBC or the triangular lattice.

In our simulations, a quench from infinite temperature is mimicked by a random configuration at $t=0$ 
that corresponds to a totally uncorrelated paramagnetic state.
Such a configuration is obtained by choosing each spin at random taking the values 
$S_i = +1 $ or $S_i = -1$ with probability a half.  We will also compare our results to the case 
in which the system is constrained such that the total magnetisation is strictly zero. 
This state is obtained by starting with all the spins $S_i = +1$ and then choosing at random $L^2/2$ among them  
to be  reversed ($L^2$ has to be even). 
The critical Ising initial states were generated by equilibrating the 
samples with a standard cluster algorithm.

Next we evolved the system at zero-temperature. At this temperature the dynamics are particularly simple. 
After choosing at random one site, the spin 
on this site is oriented along the sign of the local field (which is the sum of the nearest 
neighbour spins). If this local field is zero, the value of the spin is chosen randomly. 
$L^2$ such operations correspond to an increase of  time $\delta t = 1$. 
Quite naturally, the number of flippable spins under this rule decreases in time
and testing all the possible spins in the sample results in a waste of computer time. 
It is much faster to consider only the spins which can be actually reversed. Therefore, in order to accelerate
our numerical simulations,  we used the Continuous Time Monte Carlo (CTMC) method~\cite{Bortz}. In the following, 
the time unit is given in terms of the equivalent Monte Carlo time step. 

\vspace{0.25cm}

\begin{table}[h]
\begin{center}
\begin{tabular}{ | l ||  c | c | c | c |  c |  c |  c |}
\hline
\;  Lattice type           \;        & \ $L^M_{eq}$ \  & \ $N_M$  \ & \ $L^M_{ne}$ \ \\
 \hline 
  \; Square Lattice with PBC     \;    & \ $362$  \ &  \ $10^6$  \ & \ $8192$   \      \\
 \;  Square Lattice with FBC       \;  & \ 512 \  & \  $10^7$  \ & \  $8192$   \     \\
\;  Triangular Lattice with PBC     \;   & \ $1024$ \ & \  $10^6$  \  & \  $4096$  \  \\
\hline
\end{tabular}
\end{center}
\caption{Largest size $L^M_{eq}$ at which we equilibrated $N_M$ samples, and the largest size 
$L^N_{ne}$ ran until $10^7$ time steps and not necessarily reaching equilibrium for each type of lattice and boundary conditions.}
\label{Table1}
\end{table}

The sizes and number of configurations that we can simulate are limited by the computation time. 
For example, for PBC on the square lattice, we simulated systems until they reached a stable state for sizes up to $L=362$. 
We also simulated larger systems with a linear size up to $L=8192$ but with a maximum running time $t = 10^7$ at which the 
system is not necessarily  blocked yet.
In Table~\ref{Table1}, we specify the largest size $L^M_{eq}$ up to which we equilibrated $N_{M}$ samples. 
We also indicate the largest size $L^M_{ne}$ for which we ran the code up to a finite time $t = 10^7$. For the largest sizes, the number of 
samples $N$ have been chosen such that $N  \times L^2   \simeq 2 \ 10^{11}$. 

\section{Persistence} 
\label{sec:persistence}

In this Section we present our numerical results. 
The possible finite size dependence of the persistence probability, $P(t,L)$, is made explicit by the second argument in this function. 
In a coarsening process, the persistence 
is expected to decay algebraically, $P(t,L) \sim t^{-\theta}$,   as long as the growing length $\xi(t) \simeq t^{1/z}$ be 
shorter than the system size $L$, and it is expected to saturate to an $L$-dependent value, $P(t,L) \sim L^{-z\theta}$,
as long as $z\theta <d$ with $d$ the space dimension. The crossover between the two regimes should be captured by the scaling 
form
\begin{eqnarray}
P(t,L) \simeq L^{-z\theta} \  f\left( \frac{t}{L^z} \right) 
\qquad 
\mbox{with}
\qquad
f(x)  \sim
\left\{
\begin{array}{ll}
x^{-\theta} \qquad & x \ll 1\; , 
\\
\mbox{cst} \qquad & x \gg 1 \; . 
\end{array}
\right.
\label{eq:scaling-P}
\end{eqnarray}
For $z\theta>d$ the crossover should be pushed to infinity and the persistence
should decay to zero for all $L$~\cite{manoj_spatial_2000}. In the present case $d=2$, $z=2$ and $\theta$ will turn out to be smaller than one. Therefore, 
there  will be  a non-trivial time and size dependence of $P$ that we analyze in detail.

\subsection{Infinite temperature initial condition}
\label{subsec:Tinfty}

\begin{figure}[h]
\begin{center}
\epsfxsize=450pt\epsfysize=300pt{\epsffile{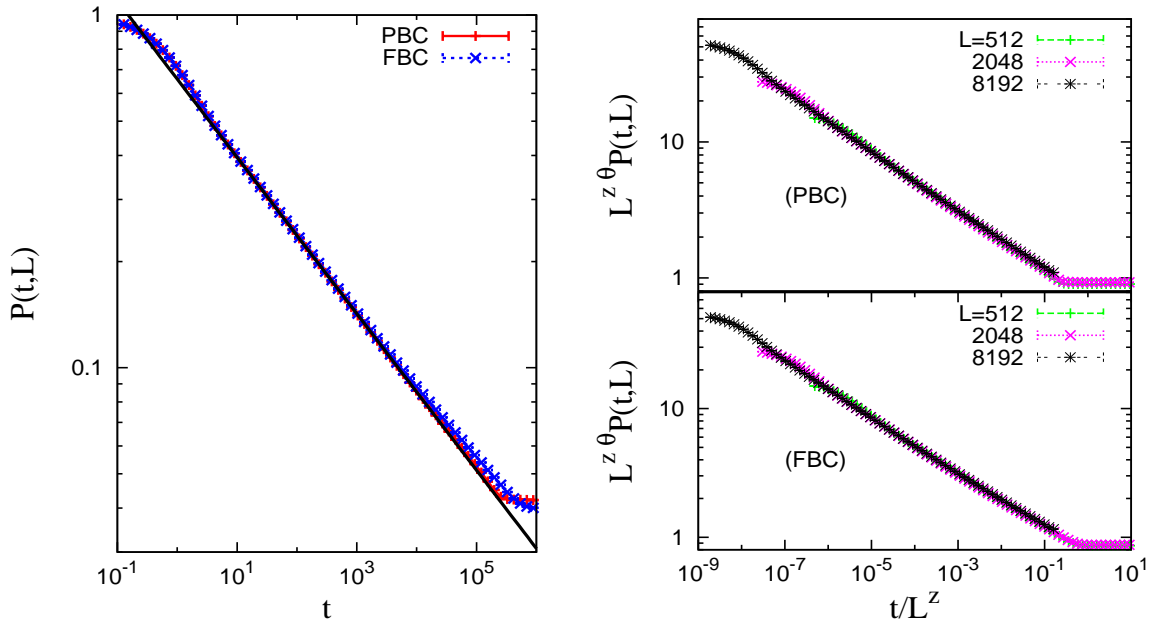}}
\end{center}
\caption{The persistence   for the ferromagnetic $2d$IM on a square lattice
with PBC and FBC (red and blue data, respectively, see the key) quenched at $t=0$ from a $1/T \to 0$ 
initial state to $T=0$. 
Main panel: raw data for $P(t,L=1024) \; vs. \; t$ in double logarithmic scale. 
The  line is the best power-law fit to the PBC data on the interval $t\in [100:10000]$ that 
yields $\theta \simeq 0.2218(4)$.
Secondary panels on the right: scaling plots for PBC (above) and FBC (below) using this value of $\theta$ and $z=2$. The 
system sizes are given in the key and the symbol (and color) code is the same as in all other figures in the paper.}
\label{Pers}
\end{figure}

In the main panel in Fig.~\ref{Pers} we show numerical results for the persistence probability as a function of  time in the ferromagnetic $2d$IM
instantaneously quenched from infinite to zero-temperature. This figure contains data for the square lattice with $L=1024$ and both FBC and PBC. 
At first sight, it seems that both cases have the  same algebraic decay for $t > 10$. 
The best fit of the data for PBC on the interval $t \in [100:10000]$ gives $\theta =0.2218 (4)$, in excellent agreement 
with previous results~\cite{DBG0,Stauffer}. This fit is shown in Fig.~\ref{Pers} as with a solid thin line. 
For longer times,  $t \gtrsim 10^6$, there is saturation of data, \textit{i.e.} $P(t,L=1024) \simeq~cst$ for 
$t \gtrsim 10^6 \simeq L^2$. 
This is expected since for $t \simeq L^2$, the 
system gets close to equilibrium and the persistence reaches a finite $L$-dependent value due to finite size effects. 
We obtain that $\lim_{t\rightarrow \infty} P(t,L) = P_{\infty}(L)  \sim L^{-2\theta}$ 
with $2\theta = 0.45 (1)$.  This is also consistent with
previous results in the literature~\cite{manoj_persistence_2000}.

The other two smaller panels on the right display the scaling plots of $P(t,L)$ according to 
 Eq.~(\ref{eq:scaling-P}). From these plots one could conclude that the scaling is very good and 
that the value of $\theta$ used is  the correct one. We will see below that this is not the case.

\subsubsection{Finite time effects}

\paragraph{Systems with PBC.}
Although the time and size dependence of the data in Fig.~\ref{Pers} seems to agree well with the power-law expectations and the 
value of the exponent $\theta$ found in the past by other authors, we would like to examine  these dependencies more carefully.
Indeed, though the fit in 
Fig.~\ref{Pers} looks good, in fact it is not. The reduced chi-squared for the power law fit 
is $\simeq 6000$ which means that it is actually a terribly bad fit. 
(The data used for the plot in Fig.~\ref{Pers} contain one million samples.)
By changing the fitting region, we observe that the value of the exponent
is changing slightly, still in the range $\theta = 0.20 - 0.225$ though always with a very large reduced chi-squared. 

\begin{figure}[h]
\begin{center}
\epsfxsize=220pt\epsfysize=150pt{\epsffile{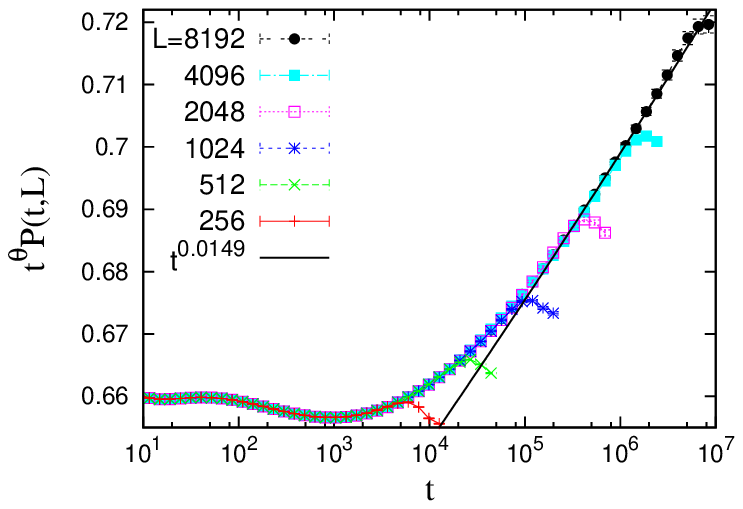}}
\epsfxsize=220pt\epsfysize=150pt{\epsffile{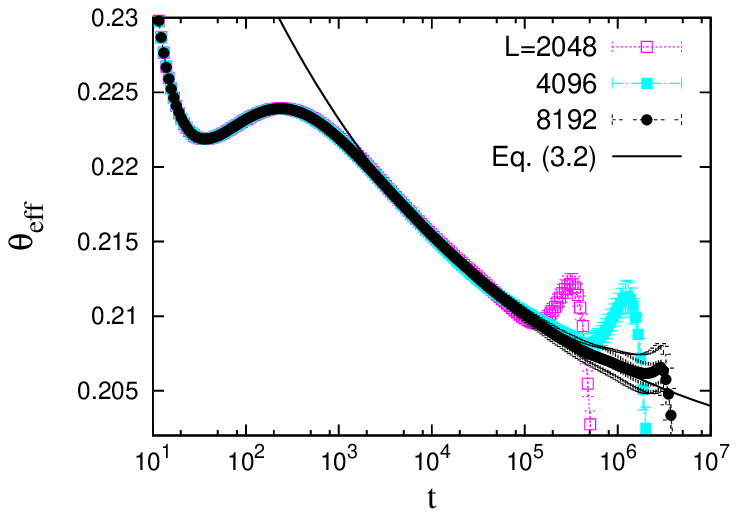}}
\end{center}
\caption{The ferromagnetic $2d$ Ising model on a square lattice
with PBC quenched from $T\to\infty$ to $T=0$ at $t=0$. The system sizes are given in the 
keys and the symbol (and color) code is the same as in all other figures in the paper.
Left panel:    $ t^\theta P(t,L) \;  vs. \; t$ with $\theta=0.2218$. The  line is a power law $t^{x}$ with $x=0.2218-0.2069=0.0149$
($0.2069$ is the value of the effective exponent in the last time-interval).
Right panel: the effective exponent $\theta_{\rm eff} \; vs. \; t$.  The line is the fit (\ref{falpha}) with parameters
$\theta_0 = 0.198 (3)$, $\theta_1 \simeq 0.07$ and  $\overline\beta \simeq 0.15$.
}
\label{Pers2}
\end{figure}

In order to improve our analysis, in the first panel of Fig.~\ref{Pers2}, we show the same data as in Fig.~\ref{Pers}
after rescaling $P(t,L)$ by the power $t^\theta$ with $\theta=0.2218$, the value obtained from the analysis 
done in Fig.~\ref{Pers}. 
For $t/L^2 \ll 1$ this quantity should be constant if the decay were well-described by this value of the exponent  $\theta$. 
This is clearly not the case. We observe different regimes as a  function of time that are characterized by different 
effective exponents $\theta_{\rm eff}(t,L)$ represented as a function of time in the right panel in the same figure.
$\theta_{\mathrm{eff}}(t,L)$ was obtained from a fit 
of the persistence probability $P(t,L)$ to a power law in the range $[t/3,3t]$.
At short  times, $ 10 \lesssim t \lesssim  100$, $\theta_{\rm eff} \simeq 0.2214$. 
In Stauffer's work~\cite{Stauffer}, measurements were done on very short time scales, 
up to $t=200$, thus in this first regime of our analysis.
Next,  for $ 100 \lesssim  t \lesssim  1000 $ the exponent increases towards $\theta_{\rm eff} \simeq 0.2241$. For still 
longer times, $t \gtrsim  100 000$, it decreases back to $\theta_{\rm eff} \simeq 0.207$. 
We also note in the left panel that the  persistence  does not depend on the size of the system
until times of the order of  $t \simeq L^2/10$. For each size, we observe a drop of $P$ beyond this time 
(this drop is followed by a rapid increase that, for clarity,  we removed from the presentation since it 
corresponds to the saturation due to the finite system size) which signals the approach to equilibrium. 
The existence of finite size effects at $t \gtrsim L^2/10$ was already observed in~\cite{Stauffer}.
\vspace{0.5cm}
\vspace{0.5cm}
From the right panel one sees that for times
$1000 \lesssim t \lesssim L^2/10$, the effective exponent 
slowly decreases with time and it is well described by a fit to the form 
\begin{equation}
\label{falpha}
\theta_{\rm eff}(t,L) = \theta_0 + \theta_1 t ^{-\overline\beta}
\end{equation}
with $\theta_0 = 0.198 (3)$, $\theta_1 \simeq 0.07$, and  $\overline\beta \simeq 0.15$ (shown with a solid line). In conclusion, we obtain the following numerical 
estimate of the persistence exponent in the ferromagnetic $2d$IM on the square lattice 
with PBC, 
\begin{equation}
\theta^{\rm PBC}_{\square} = 0.198 (3)
\; , 
\end{equation}
 in the large size and long time limits.

\begin{figure}
\begin{center}
\epsfxsize=220pt\epsfysize=150pt{\epsffile{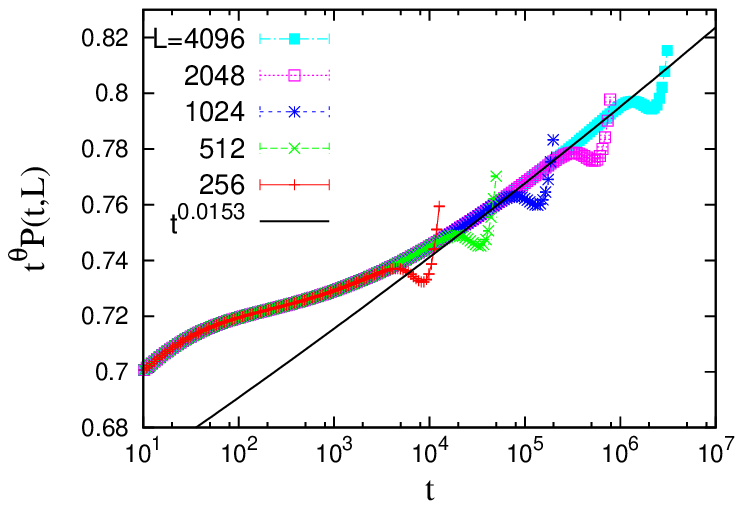}}
\epsfxsize=220pt\epsfysize=150pt{\epsffile{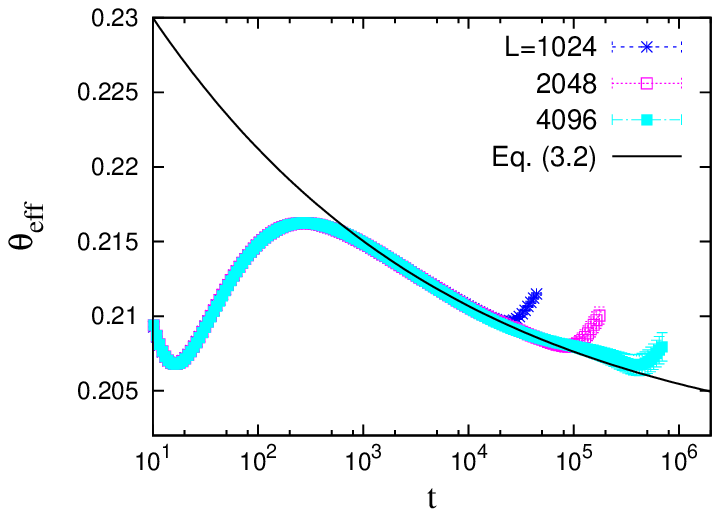}}
\end{center}
\caption{The ferromagnetic $2d$ Ising model on a triangular lattice
with PBC quenched from $T\to\infty$ to $T=0$ at $t=0$. The system sizes are given in the 
keys.
Left panel:   $t^\theta P(t,L) \; vs.  \; t$ with $\theta=0.2218$. The  line is the power-law $t^x$ with $x=0.2218-0.2065=0.0153$
($0.2065$ is the value of the effective exponent in the last time-interval).
Right panel: the effective exponent $\theta_{\rm eff} \;  vs. \; t$. The  line is the fit (\ref{falpha}) 
with parameters $\theta_0 = 0.200 (3)$, $\theta_1 \simeq 0.04$ and  $\overline\beta \simeq 0.15$. 
}
\label{Pers3}
\end{figure}

In Fig.~\ref{Pers3}, we show the same quantities for the triangular lattice with PBC. 
In the left panel we rescale the persistence with the value $\theta=0.2218$. The curves for different sizes
coincide for a given time but after this rescaling the persistence is still not constant indicating again that there are finite time effects.
We computed the effective exponent $\theta_{\rm eff}$ as explained above and we plotted it as a function of time 
for different sizes $L$ in the right panel of Fig.~\ref{Pers3}. The situation is similar to the one on the square lattice.  
To obtain a good estimate we also use a good quality fit to the form (\ref{falpha}) with $\theta_0 = 0.200 (3)$, 
$\theta_1 \simeq 0.04$, and  $\overline\beta \simeq 0.15$. 
The asymptotic value 
\begin{equation}
\theta_{\triangle}^{\rm PBC} = 0.200 (3)
\end{equation}
 is compatible with the one obtained on the square lattice $\theta_{\square}^{\rm PBC}=0.198 (3)$. We think that these
estimates are more reliable and accurate than the ones obtained with a  fit of the persistence over the whole time interval or
over just short times as done in previous studies.

\begin{figure}[t]
\begin{center}
\epsfxsize=220pt\epsfysize=150pt{\epsffile{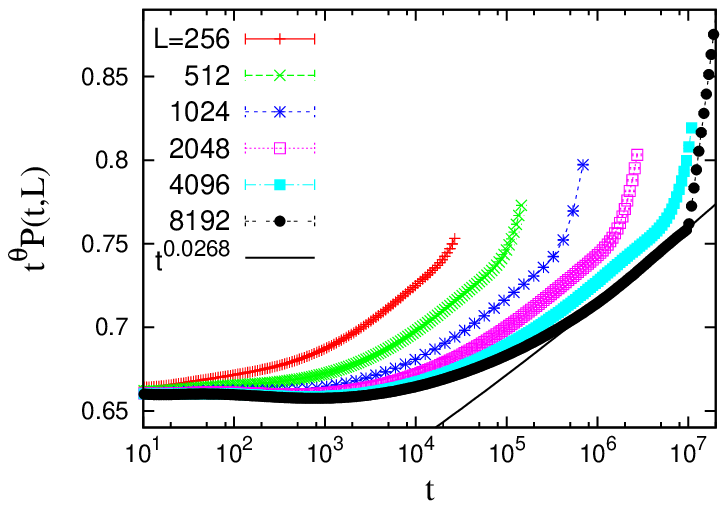}}
\epsfxsize=220pt\epsfysize=150pt{\epsffile{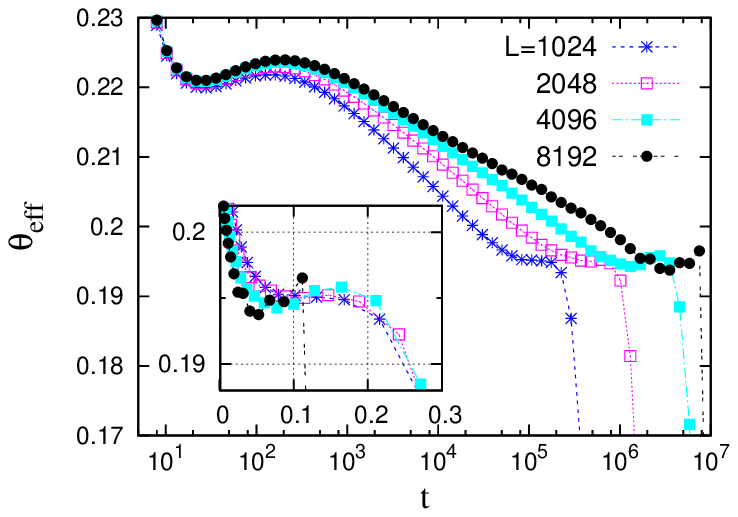}}
\end{center}
\caption{The ferromagnetic $2d$ Ising model on a square lattice
with FBC quenched from $T\to\infty$ to $T=0$ at $t=0$. The system sizes are given in the 
keys with the same symbol (and color) code as in all other figures.
Left panel:  $t^\theta P(t,L) \;   vs. \; t$ with $\theta=0.2218$. The thin line is the power-law $t^x$ with $x=0.2218-0.195=0.0268$ 
that is in better agreement with the data sets.
Right panel: the effective exponent $\theta_{\rm eff} \;  vs.  \; t$. Note the spreading of data for different $L$
and the finite length plateau at a size-dependent height close to 0.195 (see the text for a discussion). In the inset we show $\theta_{\rm eff}$ as a function 
of $t/L^2$. 
}
\label{Pers4}
\end{figure}

\paragraph{Systems with FBC.}
Now we turn to the case of FBC on the square lattice. Going back to Fig.~\ref{Pers}, we can observe that in this case
there is a deviation from scaling behaviour  for $t > 10 000$. 
A fit to a power law also gives a very large reduced chi-squared, but the fit improves as we approach 
longer times. This can be better explained in Fig.~\ref{Pers4}. 

In the left panel in Fig.~\ref{Pers4}, we plot $t^\theta P(t,L)$ 
as a function of  $t$ with $\theta=0.2218$. 
The plot does not approach a constant showing that this value of $\theta$ is not the 
definitive one. We observe that $t^\theta P(t,L)$ actually goes to a power law $t^x$ with $x \simeq 0.2218-0.195=0.0268$ for long 
times, just before reaching equilibration, suggesting that $\theta$ is close to $0.195$ in this case.
We also reckon  that $t^\theta P(t,L)$ still depends on the system size, contrary to what was observed for PBC
({\it cfr.} the left panels in Figs.~\ref{Pers2} and \ref{Pers3}). 
This fact can also be seen in the right panel of Fig.~\ref{Pers4} where we plot the effective exponent 
$\theta_{\rm eff}$ as a function of time. For each size, we observe a plateau at  times of the order of $t \simeq L^2/10$.
Thus, for FBC, finite size effects are much stronger than for PBC.
 We found that the value of $\theta$ for long times, seen as
the height of the (finite-length) plateau in the right panel figure,  
has a weak  system size dependency. 
We measured $\theta=0.1950 (1) $ for $L=256$, $\theta=0.1953 (1)$ for $L=512$ and 
$\theta=0.1954 (1)$ for $L=1024$. This last value is obtained while considering only one million samples. For large sizes, 
we have much less samples and the precision deteriorates. 
Still, it seems that a large size extrapolation will be compatible with the values obtained 
for PBC, $\theta_\square= 0.198 (3)$ or $\theta_\triangle=0.200(3)$.

\subsubsection{Finite size effects}

\paragraph{Difference between FBC and PBC.}
Let us come back to the behaviour of the effective persistence exponent
$\theta_{\rm eff}$
with the system size for PBC and FBC. Note that while for 
FBC $\theta_{\rm eff}$ has a constant value for long times and finite $L$, 
a similar result is obtained 
for PBC in the long time {\it and} large size limit only. 
When rescaling by a power of
time the PBC effective persistence exponent all the curves fall on the same master curve. Thus, for PBC, $\theta_{\rm eff}(t,L) \simeq f(t)$ with no
apparent dependence on the system size. On the contrary, for FBC, this is not the case
and one can show that $\theta_{\rm eff}(t,L) \simeq f(t/L^2)$ which is a more usual form of scaling, see the inset in Fig.~\ref{Pers4}. We have no explanation for the
intriguing difference between these two cases.

\paragraph{Saturated values.}
We now turn to the determination of the exponent $\theta$ from the final value of the persistence,
$P(t\to\infty, L) = P_{\infty}(L) \simeq L^{-2\theta}$ 
once the system is equilibrated. The exponent $2\theta$ can be determined in this case by doing a two points fit of two successive 
increasing sizes $L$. The measurement of $P_{\infty}(L)$ is in fact very time consuming  since we 
need to trully equilibrate all samples ({\it i.e.} eliminate all diagonal stripes in them). 
As a consequence, for this measurement, the largest systems that we considered are much smaller 
than for the finite time analysis. 
The largest sizes are given in Table~\ref{Table1} except for the square lattice with FBC in which case we have
 data up to $L=1024$ but with only $10^6$ samples. 

\begin{figure}[t]
\begin{center}
\epsfxsize=280pt\epsfysize=220pt{\epsffile{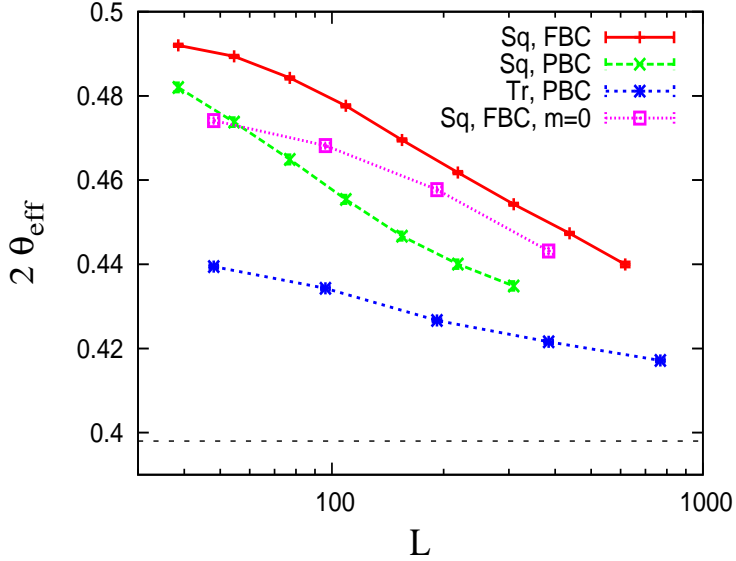}}
\end{center}
\caption{The effective exponent $2 \theta_{\rm eff}$  defined in Eq.~(\ref{eq:eff-theta}) from the saturated value of $P$ {\it vs.} $L$
in the ferromagnetic $2d$IM
with different boundary conditions and on different lattices (see the key for the symbol code) quenched from $T\to\infty$ to $T=0$ at $t=0$.}
\label{Pers5}
\end{figure}

The results for $2\theta$ are shown in Fig.~\ref{Pers5} for FBC and PBC on the square and triangular lattices. 
The effective exponent between two systems with sizes $L$  and $L'$ is 
\beq
2\theta_{\rm eff}\left(\frac{L+L'}{2}\right) = - \frac{\log \left(\displaystyle{\frac{P_\infty(L) }{P_\infty(L') }} \right) } { \log\left( \displaystyle{\frac{L}{L'} }\right) } \; .
\label{eq:eff-theta}
\eeq
We take $L'=2L$. 
In all cases, we observe that there are strong finite size corrections. 
The value of $2\theta_{\rm eff}$ decreases with size but it is hard to extrapolate an asymptotic value
from these data points.
We also show in Fig.~\ref{Pers5} data  on a triangular lattice with PBC. 
There are two advantages with this lattice. First, the diagonal crossing states are absent~\cite{BP}
and  we can therefore reach equilibrium  for relatively large sizes, up to $L=1024$. 
Second, the results for PBC have smaller finite size corrections than the ones for FBC. Moreover, we observe 
that the finite size corrections on the PBC triangular lattice are much smaller than on the PBC square lattice 
for no obvious reason. Thus, the results on the former case are much more accurate. 
A large size extrapolation of these data points gives a prediction in the range 
$2\theta_{\rm eff} = 0.40 - 0.41$ which is compatible with our previous estimates obtained from the time
evolution. However,  since the sizes are more limited, and  it is harder to
fit these data, we conclude  that this is not the optimal way to determine
$\theta$. In Fig.~\ref{Pers5} we also show the values of $2\theta_{\rm eff}$ obtained by starting from configurations with
strictly zero magnetisation. This constraint does not seem to change the behaviour. These are also subject to strong finite size corrections. 

\subsection{Critical Ising initial conditions}

\begin{figure}
\begin{center}
\epsfxsize=220pt\epsfysize=150pt{\epsffile{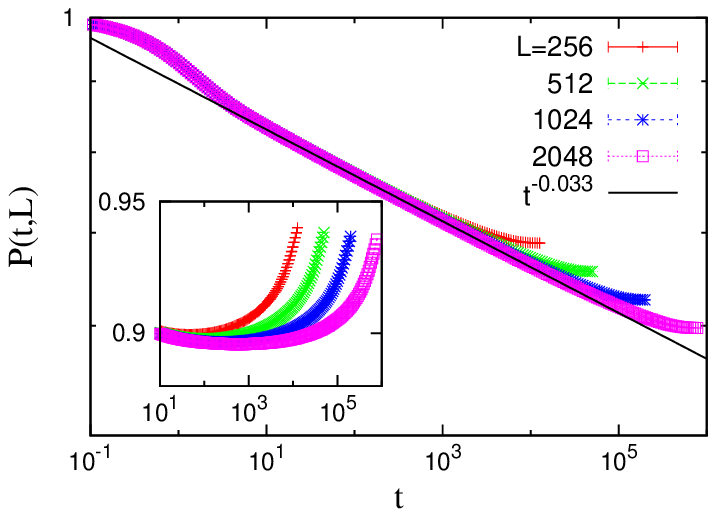}}
\epsfxsize=220pt\epsfysize=150pt{\epsffile{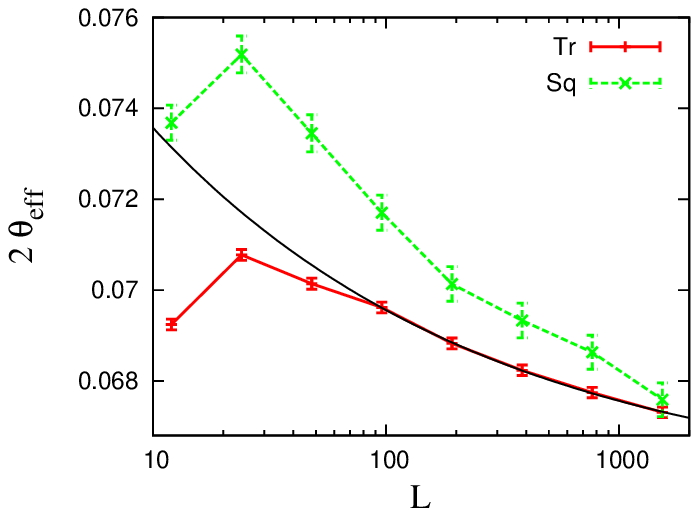}}
\end{center}
\caption{The ferromagnetic $2d$ Ising model on a triangular lattice
with PBC quenched from $T_c$ to $T=0$ at $t=0$. The system sizes are given in the 
key. Left panel: $P(t,L) \; vs.  \; t$ in double logarithmic scale. 
The solid line is a best fit to $t^{-\theta_c}$ with $\theta_c \simeq  0.033$. The inset is $t^{\theta_c} P(t,L) \; vs.  \; t$ with $\theta_c=0.033$. 
Right panel: the effective exponent $2 \theta_{\rm eff} \;  vs. \; t$ on the square and the triangular lattices with PBCs. 
The solid line is a fit of the triangular lattice data to the form $2 \theta_{\rm eff} = 2 \theta_0 + 2 \theta_1 L^{-\overline\beta}$ 
with parameters $2 \theta_0 = 0.066 (2)$, $\theta_1 \simeq 0.015$ and  $\overline\beta \simeq 0.25$. 
}
\label{Perstc}
\end{figure}

We will now concentrate on a zero-temperature quench from a critical Ising initial condition.  As for the infinite temperature
initial states, we can either follow the behaviour of the persistence with time, $P(t,L\to\infty) \simeq t^{-\theta_c}$ 
or check the size dependence of the asymptotic, 
$t\to\infty$, value, $P_\infty(L) \simeq L^{-2\theta_c}$. 
Both cases are shown in Fig.~\ref{Perstc}. The left panel shows $P(t,L)$ versus $t$ for the triangular lattice with PBC. 
As is the case for a quench from infinite temperature, 
the study of the time dependence is tricky as
the long time regime where the real persistence dominates is only attained for big systems ($L\ge 1000$).  
One measures in this way $\theta_c=0.033 (2)$ leading to the line $t^{-0.033}$ also shown in the main part of the plot.
The inset displays $t^{\theta_c} P(t,L)$ with $\theta_c=0.033$ and we see that the curves do not really 
have a very flat plateau (the upturning parts are due to the finite size saturation) and demonstrate that this 
determination of $\theta_c$ is not fully reliable.
In the right panel, we show the effective exponent $2 \theta_{\rm eff}$ versus $L$ for the square lattice (with one million samples) and 
the triangular lattice (with ten million  samples) and PBC. The solid line shows a fit of the triangular lattice data to the form $2\theta_{\rm eff} = 2 \theta_0 + 2\theta_1 L^{-\overline\beta}$
that yields $2 \theta_0 = 0.066 (2)$, $\theta_1 \simeq 0.015$ and  $\overline\beta \simeq 0.25$. 
Therefore, we estimate $\theta_c = 0.033(1)$ which is in good agreement
with the value measured from the long time limit. It is likely that the value obtained from the fit of $P_\infty(L)$ be more accurate than the one
extracted from the time dependence. 

One can derive a bound on the persistence exponent from
geometrical arguments. It was shown in~\cite{BP} that the number of spanning (for FBC) or wrapping (for PBC) clusters does not change
under a quench of the $2d$IM from its critical point up to zero-temperature.  But one can refine the analysis and 
show that, for each dynamic run, there is a one-to-one correspondence between the spanning or wrapping clusters in
the initial and final configurations (there is no coalescence nor breaking of domains)~\cite{ABCS,BCCP}. 
Thus, the spanning or wrapping clusters survive the coarsening process while all 
other clusters shrink and eventually disappear. This means that the only spins contributing to $P_\infty(L)$
have to belong to the initial spanning clusters. In the simplest case where one cluster spans both directions for a
system with FBC, the initial spanning cluster is certainly the largest one and its mass scales as $\sim L^{D}$ with
$D$ its fractal dimension. Thus, the persistence $P_\infty(L)$ must decay at least as fast or even faster than the
fraction of spins in the initial spanning cluster, $L^{D-d}$. As  $P_\infty(L)\sim L^{-2\theta_c}$,
we have $L^{D-d} \leq L^{-2\theta_c}$ and 
\begin{equation}
\label{eq:bound_pers}
2\theta_c \ge d-D
\; .
\end{equation}

Remembering that $D=d-(\beta/\nu)_{tri}$ where $(\beta/\nu)_{tri}=5/96$ are the exponents\footnote{These exponents are the same as the ones of the magnetisation of the tricritical $Q=1$ Potts model \cite{ck,SV}.} 
associated to the  size of the biggest spin cluster, one can rewrite
Eq.~(\ref{eq:bound_pers}) as $2\theta_c\ge (\beta/\nu)_{tri}$. 
The exponent that we measured complies to this inequality as
$2\theta_c = 0.066$ and $(\beta/\nu)_{tri}=5/96\simeq 0.052$. We must note that 
we have assumed that all the persistent spins in the final state belong to the initial spanning cluster.
During  evolution the persistent sites can also belong to finite size clusters, so that we cannot apply a
similar argument at finite times. 

One may rewrite the inequality (\ref{eq:bound_pers}) in another way by considering the spatial distribution of persistent sites. Several authors
have shown that the persistent sites of a system quenched from infinite temperature have a non trivial spatial
distribution~\cite{manoj_scaling_2000,manoj_spatial_2000,jain_scaling_2000,manoj_persistence_2000,ray_persistence_2004}.
Indeed, persistent sites form fractal clusters of dimension $D_p=d-z\theta$ with $z=2$, as can be easily seen from the
infinite time value of the persistence $P_\infty(L)$ which scales as $L^{-2\theta}$ so the number of
persistent sites behaves as $L^{d-2\theta}$. While we have not checked that the persistent sites have a fractal
structure in the case of a critical initial condition, they quite likely do in analogy to the previous case. We will
therefore assume that the persistent clusters have a fractal dimension $D_p=d-z\theta_c$, which is compatible with the
value of the persistence in the final state.  The inequality~(\ref{eq:bound_pers}) can then be rewritten as $D_p\le D$.
This  is quite natural since the persistent sites in the final state are a subset of the initial largest cluster and their
fractal dimension must be less than the one of the largest initial cluster.

\section{Outlook}
\label{sec:outlook}

From the time-dependent analysis of systems with PBC instantaneously quenched from infinite to zero-temperature
we estimated $\theta$ to be $\theta^{\rm PBC}_\square = 0.198(3)$ and $\theta^{\rm PBC}_\triangle=0.200(3)$. Taking the mean 
between these two values we have
\begin{equation}
\theta=0.199(2)
\; . 
\end{equation}
The analysis of systems with FBC yielded results that are compatible with this value.

The value of the persistence exponent in a zero-temperature quench from critical Ising initial conditions
that we measured,
\begin{equation}
\theta_c\simeq 0.033(1)
\; , 
\end{equation} 
is clearly different from the value measured, and computed analytically,
for a quench from infinite temperature initial conditions. The difference between
the two situations lies in the absence of ferromagnetic fluctuation long-range correlations in the infinite temperature initial conditions contrarily to the existence of 
long-range correlations of this type in the critical Ising ones. The absence of correlations
is one of the assumptions made in the analytical derivation of $\theta$, \textit{i.e.} that the initial correlation are
short-ranged so that they play no role in the universal behaviour of the dynamics~\cite{MS}. 

We have seen from the simulations that  $\theta_c \le \theta$. 
This inequality seems quite natural as one expects that less spins will flip when the initial condition is more preserved, as it is the
case for the critical initial states compared to the infinite temperature ones. One could check this idea in at least two other 
cases.

Spatial correlations can be included in the initial conditions such  $\langle s_i(0)s_{i+r}(0)\rangle\sim
1/r^{d+\sigma}$ with $\sigma=\eta-2$. We  expect that
there should be a special value $\sigma_c$ of the exponent of the initial correlations such that for correlations decaying
faster than $r^{-(d+\sigma_c)}$ we recover the short-range case.  This prediction could be tested numerically.
One could also try to extend the analysis  in~\cite{MS}, possibly combined with the one in~\cite{Bray-Humayun}, 
to obtain an analytical approximation for  $\theta$ in this case. 

Another interesting route is to study persistence after a waiting-time at zero-tem\-per\-a\-ture. 
Let us define the generalised local persistence $P(t_w,t,L)$ with $t>t_w$ as the
fraction of spins which have never flipped between time $t_w$ and  $t$. 
As the interval $[0,t]$ can be split into two disjoint intervals $[0,t_w]$ and $[t_w, t]$ one has 
$P(0,t,L)=P(0,t_w,L)P(t_w,t,L)$. Using the scaling laws for $P(0,t;L)$ and $P(0,t_w; L)$ 
one obtains that the generalised persistence should decay as:
\begin{equation}
P(t_w,t;L) = \frac{P(0,t;L)}{P(0, t_w; L)} \sim \frac{f(t/L^z)}{f(t_w,L^z)}
\; . 
\end{equation}
Choosing $t_w \ll L^z$ one has two possible limiting expressions for $P(t_w,t;L)$:
\begin{eqnarray}
P(t_w, t; L) \simeq 
\left\{
\begin{array}{l}
\left( \displaystyle{\frac{t}{t_w}} \right)^{-\theta}
\qquad
\mbox{for} 
\qquad 
t\ll L^z
\\
\left( \displaystyle{\frac{L^z}{t_w}} \right)^{-\theta}
\qquad
\mbox{for}
\qquad
t\gg L^z
\end{array}
\right.
\end{eqnarray}
Taking now $t_w \simeq L^\alpha$ with $\alpha < z$
\begin{eqnarray}
P(t_w,t;L) \simeq 
L^{-\theta (z-\alpha)}
\qquad
\mbox{for}
\qquad
t_w \simeq L^\alpha \ll L^z \ll t
\label{eq:gen_pers}
\end{eqnarray}
We checked Eq.~(\ref{eq:gen_pers}) numerically by putting the law $P(t_w, t\gg L^2; L)\sim
L^{-2\theta(\alpha)}$ to the test. The data are compatible with an exponent
$\theta(\alpha)$ that decreases linearly with $\alpha$, $\theta(\alpha)=
(1-\alpha/z)\theta(0)$, though we see a weak deviation from linearity 
for $\alpha \simeq \alpha_p$, with $\alpha_p=1/2$ in the square lattice and $\alpha_p=1/3$ on the 
triangular lattice. Indeed, 
we showed in~\cite{BCCP} that a $2d$IM ferromagnet quenched from  infinite 
temperature approaches critical percolation after a time $t_p \simeq L^{\alpha_p}$ with $\alpha_p$ taking these 
value on the two lattices. The deviation should then be due to the existence of percolating states in the system.

Finally, one could extend this analysis to the evolution at finite temperature by using, {\it e.g.},  the numerical methods proposed in~\cite{Derrida} and 
\cite{Drouffe-Godreche} to measure persistence under thermal fluctuations. A careful analysis of pre-asymptotic and finite size effects should help settling the issue 
about the dependence or independence
of the local persistence exponent on temperature in the low-temperature phase~\cite{Derrida}. One could also measure the deviations from the scaling relation $z\theta = \lambda-d+1 - \eta/2$,
with $\theta$ the global persistence exponent, $\eta$ the static anomalous and $\lambda$ the dynamic short-time exponents, expected at criticality 
beyond the Gaussian approximation~\cite{Majumdar96}.

\vspace{0.5cm}

\noindent{\large\bf Acknowledgments} 

\vspace{0.5cm}

LFC is a member of the Institut Universitaire de France.

\end{document}